\documentclass[onecolumn,10pt,nofootinbib]{revtex4}
\usepackage{graphicx,amsmath,amsthm,dcolumn,hyperref,verbatim,amsfonts,tikz,amssymb,mathtools,tikz-cd,subfig,upgreek}
\usepackage[capitalize]{cleveref}
\usepackage[symbol]{footmisc}
\usepackage{lmodern,pdfpages}
\usepackage[a4paper,margin=0.5in]{geometry}
\usepackage{tabu,ulem,tabularx,array,booktabs,float}
\usepackage[explicit]{titlesec}
\usepackage{physics}
\usepackage[toc]{appendix}

\hypersetup{colorlinks=true,urlcolor=cyan,citecolor=blue,linkcolor=red,linktocpage=true}
\usetikzlibrary{calc,automata}
\numberwithin{equation}{section}

\begin{document}
\renewcommand{\baselinestretch}{1.2}
\parskip=.5em
\title{Classical group matrix models and universal criticality}
\author{Taro Kimura}
\author{Souradeep Purkayastha}
\affiliation{Institut de Math\'ematiques de Bourgogne, Universit\'e de Bourgogne Franche-Comt\'e, 9 Avenue Alain Savary, Dijon, France}
\begin{abstract}
\section*{\large{Abstract}}
\noindent
We study generalizations of the Gross--Witten--Wadia unitary matrix model for the special orthogonal and symplectic groups. 
We show using a standard Coulomb gas treatment -- employing a path integral formalism for the ungapped phase and resolvent techniques for the gapped phase with one coupling constant -- that in the large $N$ limit, the free energy normalized modulo the square of the gauge group rank is twice the value for the unitary case. 
Using generalized Cauchy identities for character polynomials, we then demonstrate the universality of this phase transition for an arbitrary number of coupling constants by linking this model to the random partition based on the Schur measure. 
\end{abstract}
\maketitle
\tableofcontents
\newpage

\section{Introduction}
\label{sec:introduction}

Random matrix models \cite{MEHTA2004matrixmodels,Forrester:2010,Akemann:2011RMT} are often encountered in the large $N$ treatment of partition functions of gauge theories. 
In the class of unitary matrix models, one such celebrated example is the Gross--Witten--Wadia (GWW) model arising out of $\mathrm{U}(N)$ lattice gauge theory in two dimensions~\cite{gwworiginal1,wadia1980cp,wadia2012study}. 
This original model with $\mathrm{U}(N)$ as background gauge group and one non-negative real coupling constant $\beta$ demonstrated, in the statistical mechanical limit $N\rightarrow \infty$, two distinct phases demarcated by a third-order phase transition at $\beta=1$. 
The free energy $\mathcal{F}$ as a function of the coupling parameter is straightforward to compute in both phases -- $\beta\le 1$ defining the \textit{ungapped} phase, and $\beta>1$ the \textit{gapped} phase -- using functional integral and complex analytic techniques \cite{marino2015,wadia2012study}.

The GWW model was subsequently extended to a form with arbitrary coupling constants (see \cite[Sec.~4]{orlov2002tau} for a comprehensive list of related matrix models), but the phase space structure for such cases is in general not trivial to describe using the analytic methods applied for the one-parameter model. 
However we can use ideas from representation theory to demonstrate the existence of phase transitions for the general case and also describe these various phases by their free energy. 
More precisely, we may decompose the general integral in terms of Schur polynomials \cite{macdonald1998symmetric} and use the random partition related to the Schur measure \cite{okounkov2000infinite} to obtain a discrete kernel \cite{borodinokounkov1999fredholm}. For arbitrary coupling constants the phase structure shows multicritical behaviour~\cite{Periwal:1990qb,2021taroalisecond} described by the higher-order Tracy--Widom distributions~\cite{Periwal:1990qb,Le_Doussal_2018}. 

The goal of this paper is to consider generalizations of the GWW model where the gauge group $\mathrm{U}(N)$ is replaced by any of the other compact classical groups, namely, the special orthogonal groups $\mathrm{SO}(2N)$ and $\mathrm{SO}(2N+1)$, and the symplectic groups $\mathrm{Sp}(N)$. 
We are motivated to take up these specific groups as real subgroups of unitary groups. They have been studied in other physical contexts.
For example, the authors of \cite{Garcia-Garcia:2019uve} have analyzed the general problem of a specific integrable function over the compact classical groups whose integral decomposes into one over the non-trivial eigenvalues of the integrand. 
Using Andr\'eief's identity \cite{andreief1883note} they have obtained some factorization formulae for such integrals in the large $N$ limit.
It seems possible to apply these formulae to the GWW matrix model to obtain some equations linking the partition functions for the different gauge groups -- as has been done for the case of one coupling constant in \cite{Garcia-Garcia:2019uve}.

We seek to extend the analysis of the special orthogonal and symplectic GWW models paralleling the existing unitary group techniques.
First in a Coulomb gas framework we consider the ungapped phase for arbitrary coupling constants in a path integral formulation akin to the well-developed unitary treatment (see \cite{marino2015}).
For the gapped phase we switch to a resolvent formalism that converts the problem into a Riemann--Hilbert problem invoking the Plemelj formula \cite{plemelj1964problems}.
Specifically, in the case of one coupling constant we seek an explicit calculation of the free energy in parallel with the known expression in the unitary case \cite{marino2015,wadia2012study}. 

Second, we aim to demonstrate the universality of the phase transition for arbitrary coupling constants by converting the matrix integral into a sum over characters of the respective gauge group. We parallel the treatment of the unitary case in \cite{2021taroalisecond} using properties of the generalized orthogonal and symplectic Schur functions (see \cite{fultonharris} and \cite{Koike:1987JA}). 
We remark that this random partition representation can describe asymptotic behaviour of the free energy, which becomes rather involved in the Coulomb gas analysis in general. 

At this stage it is important to discuss the normalization convention chosen for the free energy in this paper. 
We choose the convention of normalizing the free energy by the square of the rank of the gauge group, which is $N$ for all the classical compact groups in their usual notation. 

We also briefly explain the terminology of splitting the free energy into the \textit{continuum} and \textit{fluctuation} components in the random partition formalism.
The continuum component is the phase-independent contribution to the free energy, which corresponds to an unrestricted sum over partitions of the Schur polynomials.
In the unitary model the continuum component is the sole contributor to the ungapped phase free energy, and we seek to see if this is true for the special orthogonal and symplectic cases as well.
The fluctuation component, on the other hand, is obtained from the restricted sum over partitions, which is expressed as the Fredholm determinant.
This part describes the universal multi-criticality of the model, and is the object of interest in the asymptotic behaviour analysis. 
\subsection*{Summary}

In \cref{sec:review} we briefly review the unitary GWW model and its phase structure in both the one-parameter and arbitrary-parameter cases.  
Beginning with the Coulomb gas model, we proceed to review the analysis in \cite{2021taroalisecond} of the continuum and fluctuation components of the free energy, which involves rewriting the partition function as a sum over Schur polynomials. 
The differing asymptotics of the fluctuation component in different regimes confirm the phase transitions.

In \cref{sec:generalformalism} we develop the special orthogonal and symplectic GWW models in the Coulomb gas formalism, following the existing treatment for the unitary case. 
For the ungapped phase -- with arbitrary couplings -- we obtain the probability distribution and free energy using variational analysis of the partition function following \cite{marino2015}. 
For the gapped phase we take up a resolvent formalism and obtain the exact free energy in the one-coupling parameter case. 
In both of these treatments, we obtain the result that the free energy for the special orthogonal and symplectic cases is twice the respective free energy in the unitary case. 
The former case, i.e. for the continuum free energy, corroborates the factorization formulae for the one-coupling case in \cite[Sec. 3.2]{Garcia-Garcia:2019uve}. 

In \cref{section:random} we develop the random partition model for the special orthogonal and symplectic cases, using the Cauchy sum formulae \cite{Garcia-Garcia:2019uve} for the orthogonal and symplectic Schur polynomials. 
This allows us to write the partition function as a sum over Schur polynomials but now with some additional multiplicative factors arising out of the Cauchy sum formulae. 
Under the large $N$ limit we find that the continuum component of the free energy is twice its unitary counterpart, but the fluctuation component is the same.
The continuum component inclusive of $\mathcal{O}\left( \frac{1}{N} \right)$ subleading contributions corroborates the result due to the Szeg\"o-Johansson theorem \cite{Johansson1997,Garcia-Garcia:2019uve}. 
The identical fluctuation component shows that the multicritical fluctuations of the classical group models are universally described by the higher-order Tracy--Widom distributions.

\section{Review of the $\mathrm{U}(N)$ model}
\label{sec:review}
We begin with an outline of the salient points of the $\mathrm{U}(N)$ GWW model. For details on the full one-parameter solution and the formal general ungapped phase solution, we refer to \cite[Sec. 8.3]{marino2015} and \cite{wadia2012study}.
For a detailed analysis of the multicritical behaviour in the general case we refer to \cite{2021taroalifirst,2021taroalisecond}.%
\footnote{
See also~\cite{beteabouttierwalsh2020multicritical} for the random partition analysis in the multicritical regime.
}
We will describe and use these techniques in greater detail in subsequent sections.
We also refer to these sections for some definitions.

The original model formulated in \cite{gwworiginal1,wadia1980cp,wadia2012study} deals with the large $N$ behaviour of the partition function
\begin{equation}
\mathcal{Z}_{\mathrm{U}(N)}(\beta)=\int_{\mathrm{U}(N)} \dd U \, \exp \left( \frac{N\beta}{2} \left(  \tr U+ \tr U^{-1} \right) \right), \label{eq:gwwU(N)original}
\end{equation}
with a coupling constant $\beta \in \mathbb{R}_{\ge 0}$. This finite $N$ integral over $\mathrm{U}(N)$ may be converted into one over its maximal torus $\mathbb{T}^{N}$, i.e. into a multiple integral over $N$ angular variables defined on $[-\pi,\pi)$. 
It is actually possible (see~\cite{BrowerRossiTan} for details; also compare with a similar result in~\cite{LeutwylerSmilga}) to write down an exact determinantal expression\footnote[2]{We are thankful to the reviewer for bringing these to our attention.} for \eqref{eq:gwwU(N)original},
\begin{equation}
\mathcal{Z}_{\mathrm{U}(N)}(\beta)= \frac{\prod_{k=0}^{N-1}k!}{(N\beta)^{N(N-1)/2}} \frac{\det\left((N\beta)^{i-1}I_{i-1}(n\beta) \right)_{i,j=1}^{N}}{\det\left((N\beta)^{i-1} \right)_{i,j=1}^{N}}, \label{eq:exactexpressionGWW1cc}   
\end{equation}
where $I_{i}$ is a modified Bessel function of the first kind parameterized by the positive integer $i$.

Under the large $N$ limit, the multiple integral can be well-approximated by a path integral over a probability distribution $\rho:[-\pi,\pi) \rightarrow \mathbb{R}_{\ge 0}$ obeying the normalization condition $\displaystyle \int_{-\pi}^{\pi} \dd{\phi} \rho(\phi) = 1$. 
The overwhelming contrbution to this path integral comes from the region around the classical saddle point, i.e. the particular probability distribution $\rho_{0}$ for which the effective action is extremized.
In the chosen coordinates this extremal probability distribution is obtained by standard variational and complex analytic methods to be

\begin{equation}
\rho_{0}(\phi)=\begin{cases}
\displaystyle
\frac{1}{2\pi} \left(1+\beta \cos \phi\right) &\beta \le 1, \\[.5em]
\displaystyle
\frac{\beta}{\pi} \cos \left(\frac{\phi}{2}\right)\sqrt{\frac{1}{\beta}-\sin^{2} \left(\frac{\phi}{2}\right)} \times \mathbf{1}_{[-\alpha,\alpha]}(\phi) & \beta > 1,
\end{cases}
\label{eq:originalgwwsolution}
\end{equation}
where in the $\beta>1$ case, $\alpha$ is defined to be the smallest positive root of $\sin \left( \frac{\phi}{2} \right)= \frac{1}{\beta}$, and the characteristic function $\mathbf{1}_{I}$ for any interval $I \subset [-\pi,\pi)$ is defined to be

\begin{align}
    \mathbf{1}_{I}(\phi) = 
    \begin{cases}
     1 & \phi \in I, \\ 0 & \phi \not\in I.
    \end{cases}
    \, 
\label{eq:characteristicfunction}
\end{align}
The phase corresponding to $\beta \le 1$ is characterized by a non-vanishing distribution over $[-\pi,\pi)$, with the exception of the transition point $\beta=1$, where, interpreting the domain as the unit circle, $\rho(-\pi)=\rho(\pi)=0$. 
There being no gaps in the distribution in this interpretation, this phase is termed the \textit{ungapped} phase. 
The distribution for phase corresponding to $\beta>1$, however, exhibits a gap over the interval $[\alpha,2\pi-\alpha]$ in the unit circle interpretation, and hence this phase is termed the \textit{gapped} phase. 
That the phase transition is of third order may be seen by observing the structure of the free energy per unit degree of freedom in the large $N$ limit, $\mathcal{F}_{\mathrm{U}}(\beta)=\lim_{N \rightarrow \infty} \frac{1}{N^{2}} \ln \mathcal{Z}_{\mathrm{U}(N)}(\beta)$:
\begin{equation}
\mathcal{F}_{\mathrm{U}}(\beta)=\begin{cases}
\displaystyle
\frac{\beta^{2}}{4} & \beta \le 1, \\
\displaystyle
\beta- \frac{1}{2}\ln \beta- \frac{3}{4} & \beta > 1.
\end{cases}
\label{eq:originalgwwfreeenergy}
\end{equation}

As an extension of this original model one can consider a version with an atmost-countable set of coupling constants $(g_{n})_{n \ge 1}$, 
\begin{equation}
\mathcal{Z}_{\mathrm{U}(N)}(g_{1},\bar{g}_{1},\ldots) \equiv \mathcal{Z}_{\mathrm{U}(N)}(\boldsymbol{\upbeta},\boldsymbol{\upgamma})=\int_{\mathrm{U}(N)} \dd U \, \exp \left( N \sum_{n \ge 1} \left( g_{n} \tr U^{n}+\overline{g}_{n} \tr U^{-n} \right) \right). \label{eq:gwworiginalmanyCC}
\end{equation}
The complex couplings -- taken as conjugates in order to have a real action -- are expressed in terms of real numbers as $g_{n}= \frac{1}{2n} \left(\beta_{n}-i \gamma_{n} \right)$, $\bar{g}_{n}= \frac{1}{2n} \left(\beta_{n}+i \gamma_{n} \right)$.
The symbols $\boldsymbol{\upbeta}$ and $\boldsymbol{\upgamma}$ represent the tuplets $(\beta_{n})_{n \ge 1}$ and $(\gamma_{n})_{n \ge 1}$ respectively. 
Using the same functional integral techniques as the one-parameter case, a formal solution to the large $N$ extremal probability distribution may be derived,
\begin{equation}
\rho(\phi)=\frac{1}{2\pi}+\frac{1}{2\pi} \sum_{n \ge 1} \left( \beta_{n} \cos n \phi+\gamma_{n}\sin n \phi \right), \label{eq:gwworiginalmanyCCrho}
\end{equation}
and the formal free energy,
\begin{equation}
\mathcal{F}_{\mathrm{U}}(\boldsymbol{\upbeta},\boldsymbol{\upgamma})= \sum_{n \ge 1} \frac{\beta_{n}^{2}+\gamma_{n}^{2}}{4n}. \label{eq:freeenergyungappedUN}    
\end{equation}
The distribution \eqref{eq:gwworiginalmanyCCrho} is non-negative -- hence a valid physical solution -- for a well-defined subset of the phase space of coupling constants around the origin $(\boldsymbol{\upbeta},\boldsymbol{\upgamma})=\boldsymbol{0}$.
For example, one may consider the Fourier series of any integrable $2\pi$-periodic, non-negative function normalized over $[-\pi,\pi)$.
In general, for an infinite set of non-zero coupling constants,~\eqref{eq:gwworiginalmanyCCrho} and \eqref{eq:freeenergyungappedUN} make sense only if the $\beta_{n}$ and $\gamma_{n}$ decay suitably with respect to $n$ for convergence. 
If required, we will assume such convergence in similar expressions appearing in the remainder of this paper.
For a~\textit{finite} set of non-zero coupling constants, these expressions are always well-defined in some region of the phase space surrounding the origin.

Further, just as for the one-parameter model, this phase is the only ungapped phase, with \eqref{eq:freeenergyungappedUN} representing the free energy in this phase. 
However, with the increasing complexity of the phase space with an increasing number of coupling constants, the exact phase-space structure and the orders of these possible phase transitions are not trivial to obtain in general using the same complex analytic techniques. 
Neither is it straightforward to obtain the functional forms of the distribution in the other possible phases. 
This necessitates the consideration of other perspectives to extract dynamical information.

\subsection{Multicritical phase space behaviour} 
\label{subsec:reviewUNrandompart}

Using connections to random partitions, the general matrix model \eqref{eq:gwworiginalmanyCC} for finite $N$ may be rephrased in terms of the Miwa variables
\begin{equation}
Ng_{n}= \frac{1}{n}t_{n}= \frac{1}{n} \tr X^{n}, \qquad 
N\bar{g}_{n}= \frac{1}{n}\bar{t}_{n}= \frac{1}{n} \tr Y^{n}, \label{eq:miwa_variablesUN}
\end{equation}
and the Cauchy sum formula over partitions restricted to depth $N$~\cite{borodinokounkov1999fredholm}
\begin{equation}
\mathcal{Z}_{\mathrm{U}(N)}(\boldsymbol{\upbeta},\boldsymbol{\upgamma})= \int_{\mathrm{U}(N)} \dd U \, \exp \left( \sum_{n=1}^{\infty} \frac{1}{n} \left( t_{n} \tr U^{n}+ \bar{t}_{n} \tr U^{-n}  \right) \right)= \sum_{\lambda| \ell(\lambda)\le N} s_{\lambda}(X)s_{\lambda}(Y). \label{eq:gwworiginalmanyCCschurexpansion}
\end{equation}
Here $X$ and $Y$ are infinite-dimensional matrices whose eigenvalues parametrize the couplings and we note that the Schur function, $s_\lambda(X)$ and $s_\lambda(Y)$, may be interpreted as a function of these eigenvalues.
See, e.g., \cite{macdonald1998symmetric} for details about the Schur functions.
The large $N$ analysis of this integral is performed in \cite{2021taroalisecond} using the techniques introduced in \cite{okounkov2000infinite,borodinokounkov1999fredholm}. 
We also have the unrestricted Cauchy sum, which is conveniently expressed using the plethystic exponential (see Appendix \ref{subsubsection:vandermonde} for definition), 
\begin{equation}
\mathcal{Z}_{\infty}(\boldsymbol{\upbeta},\boldsymbol{\upgamma}) := \lim_{N \to \infty} \mathcal{Z}_{\mathrm{U}(N)}(\boldsymbol{\upbeta},\boldsymbol{\upgamma}) =\sum_{\lambda}s_{\lambda}(X)s_{\lambda}(Y)=\operatorname{PE}\left[\tr X \tr Y \right]. \label{eq:Z_infty}
\end{equation}
The free energy\footnote[2]{Note that $X$ and $Y$ depend on $N$ through the coupling relation.} is then given by
\begin{equation}
\mathcal{F}_{\mathrm{U}}(\boldsymbol{\upbeta},\boldsymbol{\upgamma})= \lim_{N \rightarrow \infty} \frac{1}{N^{2}} \ln \mathcal{Z}_{\mathrm{U}(N)}(\boldsymbol{\upbeta},\boldsymbol{\upgamma})=\underbrace{\lim_{N \rightarrow \infty} \frac{1}{N^{2}} \ln \mathcal{Z}_{\infty}(\boldsymbol{\upbeta},\boldsymbol{\upgamma})}_{\mathcal{F}^{c}_{\mathrm{U}}(\boldsymbol{\upbeta},\boldsymbol{\upgamma})}+\underbrace{\lim_{N \rightarrow \infty} \frac{1}{N^{2}} \ln \left( \frac{\mathcal{Z}_{\mathrm{U}(N)}(\boldsymbol{\upbeta},\boldsymbol{\upgamma})}{\mathcal{Z}_{\infty}(\boldsymbol{\upbeta},\boldsymbol{\upgamma})} \right)}_{\mathcal{F}^{f}_{\mathrm{U}}(\boldsymbol{\upbeta},\boldsymbol{\upgamma})}, \label{eq:UNfreeenergy}    
\end{equation}
which may be written as the sum of the continuum contribution $\mathcal{F}^{c}_{\mathrm{U}}$ and the fluctuation contribution $\mathcal{F}^{f}_{\mathrm{U}}$. 
The continuum component does not undergo phase transitions and gives the ungapped phase free energy \eqref{eq:freeenergyungappedUN} -- this is straightforward to show from \eqref{eq:miwa_variablesUN} and \eqref{eq:Z_infty}. 
The fluctuation component is responsible for the phase transition, exhibiting different asymptotic behaviours in the ungapped and gapped phases.
Imposing the condition of real and equal couplings, i.e. $X=Y=Z$ or $\boldsymbol{\upgamma}=0$, the behaviour of the fluctuation component is analyzed. 
In the context of our work, the relevant regions of the random partition are the right and left edges, where we can analyze the contributions to the free energy.
To obtain these edge contributions, if we define the parameters
\begin{equation}
\alpha_{k}= \sum_{n=1}^{\infty} \frac{2n^{k+1}t_{n}}{k!}, \;\;\; \beta=\alpha_{0}= \sum_{n=0}^{\infty}2nt_{n}, \label{eq:redefinitionofparameters}    
\end{equation}
then clamping $\alpha_{p'}=0$ for all $p'<p$ with $p,p'$ being positive integers and $p$ being even, it is obtained that
\begin{equation}
\mathcal{F}^{f}_{\mathrm{U}}(\boldsymbol{\upbeta},\boldsymbol{0}) \sim  N^{-2} \lim_{s \rightarrow \pm \infty} \ln F_{p}(s), \;\;\; s= \frac{(\beta_{c}-\beta)N}{(\alpha_{p}N)^{\frac{1}{p+1}}}, \label{eq:freeenergyU(N)edge}
\end{equation}
where $F_{p}$ is the higher-order Tracy--Widom distribution \cite{Periwal:1990qb,Le_Doussal_2018,beteabouttierwalsh2020multicritical}\footnote{There are alternative expressions, e.g.~\cite{higherorderTWClaeysKrasovskyIts,Cafasso_2019,Akemann_2012}, for the higher-order Tracy--Widom distributions, not all of which match with \cite{Periwal:1990qb}; only those in~\cite{Le_Doussal_2018} were shown in~\cite{beteabouttierwalsh2020multicritical} to agree with the multicritical points of~\cite{Periwal:1990qb}. We thank the reviewer for pointing this out.} of order $p$, and $\beta_{c}$ has the interpretation of heralding a phase transition as $\beta$ is varied.
The order of these phase transitions may be explained from the asymmetric asymptotics of the $F_{p}$; briefly stating, one obtains the large $N$ limit edge fluctuations 

\begin{equation}
\mathcal{F}^{f}_{\mathrm{U}}(\boldsymbol{\upbeta},\boldsymbol{0}) \sim
\begin{cases}
 \mathcal{O}\left(e^{-cN} \right), & \beta < \beta_{c}, \\
 \alpha_{p}^{-2/p}|\beta_{c}-\beta|^{2(p+1)/p}+\mathcal{O}(N^{-2}), & \beta > \beta_{c}. \label{eq:freeenergyU(N)edgelargeN}
\end{cases}
\end{equation}
which again is understood in terms of the asymptotic limit, and is also modulo an additive term. 
These asymptotics follow from the those of the higher-order Tracy--Widom distributions,
\begin{equation}
F_{p}(s) \sim
\begin{cases}
 1-\mathcal{O}\left(s^{-\frac{p+1}{p}}e^{s^{\frac{p+1}{p}}} \right) & s \rightarrow \infty, \\
 \mathcal{O}\left(e^{-|s|^{\frac{2(p+1)}{p}}} \right) & s \rightarrow -\infty. \label{eq:TractWidomlargeN}
\end{cases}
\end{equation}
The factor of $c$ in the subcritical regime free energy asymptotics comes from the constants in the definition of $s$ in \eqref{eq:freeenergyU(N)edge}.
The fractional power behaviour in the supercritical regime for $p>2$ is indicative of multicriticality. 
This establishes the universality of the phase transition for the general $\mathrm{U}(N)$ model. 

Finally to conclude this review section we remark that there has been recent progress towards finding explicit multi-cut solutions to the generalized unitary GWW model -- see e.g.~\cite[Sec. 4]{Oota_2022} for details. 

\section{Coulomb gas formalism for the $\mathrm{SO}$ \& $\mathrm{Sp}$ models}
\label{sec:generalformalism}

We now broach the main subject of this paper, the generalization of \eqref{eq:gwworiginalmanyCC} to the special orthogonal groups $\mathrm{SO}(2N)$ and $\mathrm{SO}(2N+1)$, and the symplectic groups $\mathrm{Sp}(N)$.
We adapt the variational formalism from that of the original $\mathrm{U}(N)$ model in \cite[Sec. 8.3]{marino2015}, to develop the formal solution in the ungapped regime in \cref{subsection:ungappedphase}.
For the analysis of the gapped regime with one coupling constant, we use a slightly different formalism (following \cite{eynardkimuraribault2015}) in \cref{subsection:gappedphase}.

We begin with the important observation that we consider the compact real forms of these groups, i.e. the maximal compact subgroups of their complexified forms.
That is, $\mathrm{SO}(N) \subset \mathrm{SO}(N,\mathbb{C})$ is composed of $N \times N$ real special orthogonal matrices, and $\mathrm{Sp}(N)=\mathrm{Sp}(2N,\mathbb{C}) \cap \mathrm{U}(2N)$ is composed of $2N \times 2N$ complex symplectic and unitary matrices.
For these groups, $\tr X=\tr X^{-1} \in \mathbb{R}$ for any $X \in \mathrm{SO}(N),\mathrm{Sp}(N)$, and hence to get a real action, the generalized GWW model must have real coupling constants -- in fact, it may be parametrized by just one set of real coupling constants $\left( g_{n} \right)_{n \ge 1}$ with $g_{n}=\frac{\beta_{n}}{n}$:
\begin{equation}
\mathcal{Z}_{\mathrm{G}(N)}(\boldsymbol{\upbeta})=\int_{\mathrm{G}(N)} \dd X \, \exp \left( N \sum_{n\ge 1} g_{n} \tr  X^{n} \right), \label{eq:gwwgeneralizedmanyCC}
\end{equation}
where $\mathrm{G}(N)$ is any of the groups
\begin{equation}
\mathrm{G}(N)=\mathrm{SO}(2N),\, \mathrm{SO}(2N+1),\,\mathrm{Sp}(N). \label{eq:listofgaugegroups}
\end{equation}

Under the large $N$ limit, matrix integrals \eqref{eq:gwwgeneralizedmanyCC} become path integrals over an appropriate functional measure, and are computed by switching from an integral over $G$ to one over its maximal torus using the Weyl integral formula (see Appendix~\ref{subsection:weylintegration}).
The groups \eqref{eq:listofgaugegroups} are of rank $N$, i.e. the maximal torus is of $N$ dimensions in all cases and can be parametrized by $N$ angular variables $\phi_{i} \in [-\pi,\pi)$, $1\le i \le N$.
The change of variables introduces a measure factor, the generalized Vandermonde determinant.
We write down this factor for each of the groups \eqref{eq:listofgaugegroups} in Appendix \ref{subsubsection:vandermonde}. 

Employing the Weyl integral formula and ignoring numerical prefactors\footnote{The Weyl group cardinality prefactor and other terms from the measure change can be thought of as being absorbed into the eventual path integral. Upon doing so, they enter the exponential as $\mathcal{O}(N)$ terms and hence may be ignored -- see the discussion following \eqref{eq:gwwgeneralizedmanyCCgeneralformlargeN}.}, in all these cases we may write the finite $N$ partition function \eqref{eq:gwwgeneralizedmanyCC} as

\begin{equation}
\mathcal{Z}_{\mathrm{G}(N)}(\boldsymbol{\upbeta})= \int_{-\pi}^{\pi} \prod_{j=1}^{N} \frac{\dd \phi_{j}}{2 \pi} \exp \left( \sum_{\substack{k,l=1 \\ k \ne l}}^{N} \Delta(\phi_{k},\phi_{l})+ \sum_{k=1}^{N} \Xi(\phi_{k}) +N \sum_{k=1}^{N} V (\phi_{k})\right).  \label{eq:gwwgeneralizedmanyCCgeneralform}
\end{equation}
Here $\Delta(\phi_{k},\phi_{l})$ and $\Xi(\phi_{k})$ come from the Vandermonde determinant, and the \textit{single-variable action} $V(\phi_{i})$ comes from the switch to the maximal torus. The specific forms of $\Delta$ and $\Xi$ for the groups \eqref{eq:listofgaugegroups} are listed in Appendix \ref{subsubsection:fredholm}, and the respective single-variable actions are listed in Appendix \ref{subsubsection:singlevariableaction}. We define the effective action $V_\textrm{eff}(\phi_{1},\ldots,\phi_{N})$ as
\begin{equation}
N^{2}V_{\mathrm{eff}}(\phi_{i}) = \sum_{\substack{k,l=1 \\ k \ne l}}^{N} \Delta(\phi_{k},\phi_{l})+\sum_{k=1}^{N} \Xi(\phi_{k})+N \sum_{k=1}^{N} V(\phi_{k}), \label{eq:disceteNactions}
\end{equation}  
and the free energy,
\begin{equation}
\mathcal{F}_{\mathrm{G}}(\boldsymbol{\upbeta})=\lim_{N \rightarrow \infty}\frac{1}{N^{2}} \ln \mathcal{Z}_{\mathrm{G}(N)}(\boldsymbol{\upbeta}). \label{eq:freeenergygeneralclassdefn} 
\end{equation}

\subsection{Ungapped phase: formal solution}
\label{subsection:ungappedphase}

Under the large $N$ limit the partition function \eqref{eq:gwwgeneralizedmanyCCgeneralform} becomes a path integral over a normalized distribution of eigenvalues $\rho$ defined on $[-\pi,\pi)$,

\begin{equation}
\mathcal{Z}_{\mathrm{G}(N)}(\boldsymbol{\upbeta})[\rho] = \int_{\|\rho\|_{1}=1} \mathcal{D} \rho \, \exp \left( N^{2} \underbrace{ \left(  \mathcal{P} \int_{-\pi}^{\pi} \int_{-\pi}^{\pi} \dd \phi \dd \varphi \, \Delta(\phi,\varphi)\rho(\phi) \rho(\varphi)+ \frac{1}{N} \mathcal{P} \int_{-\pi}^{\pi} \dd \phi \, \Xi(\phi)+ \int_{-\pi}^{\pi} \dd \phi \, V(\phi) \rho(\phi) \right)}_{S_{\textrm{eff}}[\rho]} \right), \label{eq:gwwgeneralizedmanyCCgeneralformlargeN} 
\end{equation}
where we define the large $N$ effective action $S_{\mathrm{eff}}[\rho] = \lim_{N \rightarrow \infty} V_{\mathrm{eff}}(\phi)$. The notation $\|f\|_{1}$ denotes the $L^{1}$-norm of a function $f$ defined on $[-\pi,\pi)$.
We remark that double integral over $\phi$ and $\varphi$ is to be understood at the Cauchy principal value as $\Delta$ is singular along the diagonals $\phi=\pm \varphi$; hence the diagonals are understood to be removed from the integration. 
We also note that the term containing $\Xi$ -- similarly possibly singular, see \eqref{eq:xi} -- is of order $N$ and can be ignored in subsequent calculations, and a similar truncation can be done to the single-variable action in the $\mathrm{SO}(2N+1)$ case (see Appendix \ref{subsection:maximaltori}). 
For the three cases, we find identical expressions for $\Delta$ upto $\mathcal{O}(N^{2})$, and hence, identical effective actions.

The integral \eqref{eq:gwwgeneralizedmanyCCgeneralformlargeN} may be well-approximated at the stationary, i.e. classical configuration $\rho_{0}$ which extremizes $S_{\mathrm{eff}}[\rho]$. Consequently the path integral analogue of the discrete free energy defined earlier\footnote[3] {We consider the two to be equal, though from a rigorous mathematical standpoint one may point out that convergence of functional integrals is not well-defined in general.}, $\mathcal{F}_{\mathrm{G}}(\boldsymbol{\upbeta})[\rho]=\lim_{N \rightarrow \infty} \frac{1}{N^{2}}\ln \mathcal{Z}_{\mathrm{G}(N)}[\rho]$, may be approximated as $\mathcal{F}_{\mathrm{G}}=\mathcal{F}_{\mathrm{G}}(\boldsymbol{\upbeta})[\rho_{0}] \approx S_{\mathrm{eff}}[\rho_{0}]$. The extremal distribution $\rho_{0}$ may be obtained using the method of Lagrange multipliers. We seek to extremize the free energy density

\begin{equation}
\mathcal{F}_{\mathrm{G}}(\boldsymbol{\upbeta})[\rho,\xi]=S_{\mathrm{eff}}[\rho]+\xi \left(\int_{-\pi}^{\pi} \dd \phi \, \rho(\phi)-1 \right), \label{eq:freeenergygeneral}
\end{equation}
where $\xi$ accounts for the normalization condition. Noting that $\Delta$ is symmetric in its arguments in all cases, taking functional derivatives with respect to $\rho$ and differentiating the resultant equation with respect to $\phi$ to eliminate $\xi$ yields

\begin{subequations}
\label{eq:variation}
\begin{equation}
2 \mathcal{P} \int_{-\pi}^{\pi}\dd \varphi \, \Delta(\phi,\varphi) \rho(\varphi)+V(\phi)+\xi=0, \label{eq:variation1}
\end{equation}
\begin{equation}
2 \mathcal{P} \int_{-\pi}^{\pi}\dd \varphi \, \frac{\partial\Delta}{\partial \phi}(\phi,\varphi) \rho(\varphi)+V'(\phi)=0. \label{eq:variation2}
\end{equation}
\end{subequations}  
These are Fredholm integral equations of the first kind, and we again note that the integrals in \eqref{eq:variation} are understoood to be at the Cauchy principal value as the respective kernels $\Delta$ and $\frac{\partial \Delta}{\partial \phi}$ are singular along the diagonals $\phi=\pm \varphi$. The solutions to \eqref{eq:variation1} and \eqref{eq:variation2}, when plugged back in \eqref{eq:freeenergygeneral}, give us the approximate free energy $\mathcal{F}_{\mathrm{G}}$. This calculation simplifies using \eqref{eq:variation1}:
\begin{eqnarray}
\mathcal{F}_{\mathrm{G}}(\boldsymbol{\upbeta}) = \mathcal{P} \int_{-\pi}^{\pi} \dd \phi \, \left[ \int_{-\pi}^{\pi} \dd \varphi \, \Delta(\phi,\varphi) \rho_{0}(\varphi)+V(\phi) \right] \rho_{0}(\phi)= \frac{1}{2} \int_{-\pi}^{\pi} \dd \phi \, V(\phi) \rho_{0}(\phi)- \frac{\xi}{2}. \label{eq:simplifiedfreeenergy} 
\end{eqnarray}

\subsubsection*{Solution for arbitrary coupling constants}

Using the common first-order derivative of the integral equation kernel \eqref{eq:fredholm234} and single-variable action \eqref{eq:single_variable_actionSO(2N),Sp(N)} for the three cases, we proceed to obtain the formal solution for the ungapped phase for the model \eqref{eq:gwwgeneralizedmanyCC}. The integral equation \eqref{eq:variation2} becomes 
\begin{equation}
\mathcal{P} \int_{-\pi}^{\pi} \dd \varphi \, \left[ \cot \left( \frac{\phi+\varphi}{2} \right)+\cot \left( \frac{\phi-\varphi}{2} \right) \right] \rho(\varphi)=2 \sum_{n \ge 1} \beta_{n} \sin n \phi. \label{eq:fredholmequationungapped}
\end{equation}
The symmetry $\varphi \leftrightarrow -\varphi$ of the kernel implies that $\rho$ is even. We assume an even Fourier series for $\rho$,
\begin{equation}
\rho(\varphi)=\frac{1}{2\pi}+\sum_{n=1}^{\infty} B_{n} \cos n \varphi, \label{eq:rhofourier}
\end{equation} 
with Fourier coefficients $(B_{n})_{n \ge 1}$, and the constant term fixed by the normalization, and decompose the integral equation kernel by using the Fourier series for $\cot\left( \frac{x}{2} \right)$ (see Appendix \ref{subsubsec:fourierseries}),
\begin{equation}
\cot \left( \frac{\phi+\varphi}{2} \right)+\cot \left( \frac{\phi-\varphi}{2} \right) =4 \sum_{n=1}^{\infty} \sin n \phi \cos n \varphi. \label{eq:fredholmfourier}
\end{equation}
Plugging in \eqref{eq:rhofourier} and \eqref{eq:fredholmfourier} into \eqref{eq:fredholmequationungapped} and using standard Fourier series techniques yields $2\pi B_{n}= \beta_{n}$ for all defined $\beta_{n}$, and vanishing otherwise.
Hence we obtain the distribution for the ungapped phase,
\begin{equation}
\rho_{0}(\phi)= \frac{1}{2\pi}+\frac{1}{2\pi} \sum_{n \ge 1} \beta_{n} \cos n \phi, \label{eq:ungappedsolutionrho}
\end{equation}
which is of the same form as the $\mathrm{U}(N)$ solution \eqref{eq:gwworiginalmanyCCrho}, but without the imaginary coupling constants. 

Using \eqref{eq:simplifiedfreeenergy} we now calculate the free energy in this phase. The extremal value of $\xi$ can now be found out from \eqref{eq:variation1}, with the trick of setting $\phi=0$ without loss of generality to simplify the calculation. The solution now proceeds exactly as in the $\mathrm{U}(N)$ case \cite[Ch. 8.3]{marino2015} except for a factor of $2$, giving $\xi=0$, and the free energy density is calculated by the usual Fourier series integrals,
\begin{equation}
\mathcal{F}_{\mathrm{G}}(\boldsymbol{\upbeta})=\frac{1}{2} \int_{-\pi}^{\pi} \dd \phi \, V(\phi) \rho_{0}(\phi)= \sum_{n \ge 1} \frac{\beta_{n}^{2}}{2n}. \label{eq:freeenergy}
\end{equation}
The free energy density in the ungapped phase noticeably differs by a factor of $2$ from that obtained for the $\mathrm{U}(N)$ case \eqref{eq:freeenergyungappedUN}, with the imaginary couplings switched off, i.e. $\mathcal{F}_{\mathrm{G}}(\boldsymbol{\upbeta})=2 \mathcal{F}_{\mathrm{U}}(\boldsymbol{\upbeta},\boldsymbol{0})$. This may be viewed as a consequence of the $2N$ dimensions in consideration rather than $N$. 
We also note that this is consistent with the identities derived for the one-coupling constant partition function in \cite[Sec. 3.2]{Garcia-Garcia:2019uve}, which relate the $\mathrm{U}(2N)$ and $\mathrm{U}(2N+1)$ partition functions to the $\mathrm{G}(N)$ ones through some product formulae valid at large $N$.
The limit is taken before the normalization, showing that this corresponds to the continuum free energy.
We remind ourselves that, when comparing the continuum free energy from these formulae in~\cite{Garcia-Garcia:2019uve}, the normalization factor at leading order for the unitary partition functions would be $\frac{1}{4N^{2}}$ in our chosen convention. 

\subsection{Gapped phase 
}
\label{subsection:gappedphase}

To obtain the gapped phase solutions we use a technique employing a resolvent (see \cite[Sec. 3]{eynardkimuraribault2015} for details) which is different from the previous presentation. 
We consider the finite $N$ effective action \eqref{eq:disceteNactions} and take its derivatives with respect to the angular parameters, getting the $N$ equations
\begin{equation}
-V'(\phi_{k})=  \frac{2}{N}\sum_{\substack{l=1 \\ l \ne k}}^{N} \frac{\partial \Delta}{\partial \phi_{k}}(\phi_{k},\phi_{l})+\frac{1}{N} \frac{\partial \Xi}{\partial \phi_{k}}(\phi_{k}). \label{eq:finiteNeffactionderivative}
\end{equation}
To pass to the large $N$ limit we introduce the resolvent for the integral equation kernel \eqref{eq:fredholm234},
\begin{equation}
W(\phi)= \frac{2}{N} \sum_{l=1}^{N} \frac{\partial \Delta}{\partial \phi}(\phi,\phi_{l})= \frac{1}{N} \sum_{l=1}^{N} \left[ \cot \left( \frac{\phi+\phi_{l}}{2} \right)+\cot \left( \frac{\phi-\phi_{l}}{2} \right) \right], \label{eq:resolvent}
\end{equation}
which has $2N$ poles $\phi=\pm \phi_{l}$ in general and is an odd function in $\phi$. 
Then, the saddle point equation \eqref{eq:finiteNeffactionderivative} for these kernels can be rewritten as
\begin{equation}
-V'(\phi_{k})=  \frac{1}{2} \left[W(\phi_{k}+i0)+W(\phi_{k}-i0) \right]+\frac{1}{N} \frac{\partial \Xi}{\partial \phi_{k}}(\phi_{k}). \label{eq:finiteNeffactionderivativemodified}
\end{equation}
We use the notation $f(z\pm i0)=\lim_{\epsilon \rightarrow 0^{+}}f(z\pm i\epsilon)$ for a complex function $f$, assuming such a limit is defined. 

Upon taking \eqref{eq:finiteNeffactionderivativemodified} to the large $N$ limit, the $\Xi$-term vanishes, the poles of the resolvent \eqref{eq:resolvent} are promoted to a cut singularity $\mathcal{C}$ in the complex plane, and writing $V$ explicitly from \eqref{eq:single_variable_actionSO(2N),Sp(N)}, \eqref{eq:finiteNeffactionderivativemodified} transforms into the crossing-cut equation
\begin{equation}
2 \sum_{n \ge 1} \beta_{n} \sin n \phi=  \frac{1}{2} \left[W(\phi+i0)+W(\phi-i0) \right], \label{eq:largeNresolventequation}
\end{equation}
which we need to solve for $W(\phi)$. We note that the complex asymptotic behaviour of the finite $N$ resolvent \eqref{eq:resolvent} results in the following asymptotics for the large $N$ resolvent, defining $z=i\phi$:
\begin{equation}
W(\phi) \xrightarrow{z \rightarrow \pm \infty} \pm 2i. \label{eq:resolventpoles}
\end{equation}
We remark that the domain of consideration may be promoted to a Riemann surface $\mathcal{R}$ -- for example, if there is one cut in the principal domain, we may promote to the Riemman sphere; for multiple cuts we shall have higher-genus Riemann surfaces. 
We work in the usual complex coordinates interpreting $\mathbb{C}$ as a subset of $\mathcal{R}$. 
To recover the density function, we can assume that $\rho$ will be even as with in the ungapped phase, and use the relation 
\begin{equation}
\rho(\phi)= \frac{1}{8\pi i} \left[ W(\phi- i0)-W(\phi+ i0) \right], \label{eq:densityrecoveryresolvent}
\end{equation}
which may be derived by contour integration arguments using the content of the Sokhotski--Plemelj theorem \cite{sokhotski1873definite,plemelj1964problems}. A sketch of the argument for this result at finite $N$ level may be found in Appendix \ref{subsubsection:plemelj}. Now we define the following function $f: \mathcal{R} \rightarrow \mathcal{R}$ which is given in our local complex coordinates by
\begin{equation}
f(\phi)= - \left[ 2 V'(\phi) + W(\phi) \right]W(\phi) = \left[4 \sum_{n \ge 1} \beta_{n} \sin n \phi -W(\phi) \right]W(\phi), \label{eq:definef}
\end{equation}
and using \eqref{eq:largeNresolventequation} it may be shown that $f$ is regular, i.e. $f(\phi+i0)=f(\phi-i0)$ on the real line. From \eqref{eq:resolventpoles} we get
\begin{equation}
f(\phi) \xrightarrow{z \rightarrow \pm \infty} 4 \beta_{\bar{n}}e^{\pm \bar{n} z}, \label{eq:asymptoticsf}
\end{equation}
where $\bar{n}$ is the largest $n$ such that $\beta_{n} \ne 0$. 
In the complex plane these asymptotics fix the function $f$ to be a polynomial in $e^{\pm z}$ of degree $\bar{n}$ with only the coefficients of $e^{\pm \bar{n}z}$ fixed,
\begin{equation}
f(\phi)= 8 \beta_{\bar{n}} \cos \bar{n} \phi + 4c + 4\sum_{n = 1}^{\bar{n}-1} \qty( a_n \cos n \phi + b_n \sin n \phi ), \label{eq:formoffsemifixed}
\end{equation}
with unknown coefficients\footnote[1]{The situation of $\bar{n}=\infty$, i.e. infinite coupling constants, is more subtle in the Coulomb gas formalism. At present for the purposes of this section we restrict $\bar{n}$ to be finite.}
$(a_{n},b_{n})_{1 \le n \le \bar{n}-1}$ and $c$.
 
From \eqref{eq:definef}, the resolvent may be provisionally written as
\begin{equation}
W(\phi)
=-V'(\phi) - \sqrt{V'(\phi)^2 - f(\phi)} 
=2 \sum_{n \ge 1} \beta_{n} \sin n \phi - 2 \sqrt{\left(\sum_{n = 1}^{\bar{n}} \beta_{n} \sin n \phi \right)^{2}- \frac{f(\phi)}{4} 
}. \label{eq:resolventderived}
\end{equation}
The sign factor in front of the square root is determined to be consistent with the asymptotic behavior of the resolvent \eqref{eq:resolventpoles}.
Since $V'(\phi)^2$ and $f(\phi)$ are degree $2\bar{n}$ and $\bar{n}$ trigonometric functions, we may write
\begin{align}
    V'(\phi)^2 - f(\phi) = A(\phi)^2 B(\phi) \, ,
\end{align}
where $A(\phi)$ is a function of degree $(\bar{n} - m)$ for a certain integer $m \le \bar{n}$ and $B(\phi)$ is a degree $2m$ function,
\begin{align}
    B(\phi) = \prod_{\alpha = 1}^{2m} \sin (\phi - \phi_\alpha)
    \, .
\end{align}
We assume the ordering of the parameters $-\pi \le \phi_1 < \phi_2 < \cdots < \phi_{2m} \le + \pi$.
Then, we obtain the density function from the discontinuity of the resolvent~\eqref{eq:densityrecoveryresolvent} as follows,
\begin{align}
    \rho(\phi) = \frac{1}{4\pi} A(\phi) \sqrt{-B(\phi)} \times \mathbf{1}_{\mathcal{C}}(\phi) 
    \, ,
    \label{eq:probabilityfoundresolvent}
\end{align}
which has a support on $\mathcal{C} = \bigsqcup_{\alpha = 1}^m [\phi_{2\alpha-1}, \phi_{2\alpha} ]$, so that this is called the $m$-cut solution.

\subsubsection*{Solution with one coupling constant}

The density function \eqref{eq:probabilityfoundresolvent} may be solved for the general case of an arbitrary number of coupling constants (we refer to \cite[Sec. 3.2.2]{eynardkimuraribault2015} for details; also see\footnote{We thank the reviewer for bringing these to our attention.} \cite{JURKIEWICZ1990178,Bonnet_2000} for corresponding literature in Hermitian matrix models for a comparative understanding, and \cite{DeiftKriecherbauerMcLaughlin} for a treatment involving Jacobi theta functions).
For general $\bar{n}$ the solution is rather involved, but the case of $\bar{n}=1$, i.e. one coupling constant $\beta_{l}=\delta_{l,1}\beta$ is tractable, for then we have just one undetermined coefficient $c$ which may be found by the normalization condition. 
In this case it is clear that for $\beta > 1$ we expect a gap to appear symmetric around $\pm \pi$.
From \eqref{eq:probabilityfoundresolvent} it may be shown that we get a distribution with exactly one gap, i.e. $\mathcal{C}=[-\alpha,\alpha]$ in the principal interval, as a physically expected continuation only if $-2\beta < c<1+\beta^{2}$. 
Further, it turns out that $c=2\beta$ uniquely fixes the normalization (see Appendix \ref{subsubsec:proofofc}), and so from \eqref{eq:probabilityfoundresolvent} we get the gapped regime distribution

\begin{equation}
\rho_{0}(\phi)= \frac{\beta}{\pi} \cos \left(\frac{\phi}{2}\right)\sqrt{\frac{1}{\beta}-\sin^{2} \left(\frac{\phi}{2}\right)}, \label{eq:distributiongappedphaseoneconstant}
\end{equation} 
with domain $[-\alpha,\alpha]$ where $\sin \left( \frac{\alpha}{2} \right)= \frac{1}{\beta}$. Hence we have shown that for the case of one real coupling constant, the orthogonal and symplectic cases have a solution identical in structure to the solution for the $\mathrm{U}(N)$ model \eqref{eq:gwwU(N)original}. 

The free energy density in this phase may now be calculated using \eqref{eq:simplifiedfreeenergy}. The calculation is identical in structure with that for the $\mathrm{U}(N)$ case in \cite{marino2015}, except for the factor of $2$ from the single-variable action, which arises from the consideration of $2N$ dimensions rather than $N$. 
We get 
\begin{equation}
\mathcal{F}_{\mathrm{G}}(\beta)=2\beta-\ln \beta- \frac{3}{2}. \label{eq:freeenergygappedphaseoneCC}
\end{equation}
A comparison with \eqref{eq:freeenergy} for one coupling constant shows that the phase transition at $\beta=1$ is third order, and one with \eqref{eq:originalgwwfreeenergy} gives $\mathcal{F}_{\mathrm{G}}(\beta)=2\mathcal{F}_{\mathrm{U}}(\beta)$.

\section{Random partition formulation}
\label{section:random}

The variational and resolvent techniques described so far output a reasonable amount of qualitative information about the generalized GWW model. 
In particular for the ungapped phase we can fully describe the system, and for the one-constant case, also describe the gapped phase and demonstrate the third-order nature of the phase transition. 
However as we have seen this is not trivial to do for arbitrary coupling constants.

The universality of the phase transition may be demonstrated using character theory properties of the relevant groups.
The irreducible characters $\mathrm{U}(N)$ and the groups \eqref{eq:listofgaugegroups} can be indexed using partitions of positive integers \cite{fultonharris}. 
For partitions of length less than or equal to the dimensions of the maximal torus, these characters are respectively equivalent to the generalized Schur functions $s_{\lambda},o_{\lambda}^{\operatorname{even}},o_{\lambda}^{\operatorname{odd}},sp_{\lambda}$. 
The definitions of these Schur, orthogonal Schur and symplectic Schur functions may be found in \cite{fultonharris}.
The arguments of these functions are tuplets of parameters -- possibly infinite. 
These may be conveniently interpreted as the eigenvalues of (possibly infinite dimensional) matrices. 
For matrices $X,Y$ with eigenvalues $(x_{i})_{i \ge 1},(y_{i})_{i \ge 1}$ respectively, and partition $\lambda$, these generalized Schur functions obey the Cauchy sum formulae \cite{Koike:1987JA,Garcia-Garcia:2019uve}\footnote[1]{For the arguments of the functions, we group all the eigenvalues under the label $x$. In the literature it is typically $(x,x^{-1})$ in the $\mathrm{Sp}(N)$ and $\mathrm{SO}(2N)$ cases, and $(1,x,x^{-1})$ in the $\mathrm{SO}(2N+1)$ case. Also note that the eigenspectrum in each case is equivalent to that of a corresponding maximal torus (see Appendix \ref{subsection:maximaltori}), because any element of a Lie group is conjugate to some element of any maximal torus.}
\begin{subequations} \label{eq:Cauchy_sum}
\begin{align} 
    \mathrm{U}(N):&&  \sum_{\lambda} s_\lambda (X) s_\lambda(Y) 
    & = \prod_{i,j} (1 - x_i y_j)^{-1} 
    = \operatorname{PE} \left[\tr X \tr Y \right], \label{eq:Cauchy_sum_U}\\
    \mathrm{SO}(2N):&& \sum_{\lambda} o^{\operatorname{even}}_\lambda (X) s_\lambda(Y) 
    & = \prod_{i,j} (1 - x_i y_j)^{-1} \prod_{i \le j} (1 - y_i y_j) \nonumber \\ && &
    = \operatorname{PE} \left[\tr X \tr Y \right]
    \operatorname{PE} \left[ \frac{1}{2} \left( -\tr Y^{2}-\left(\tr Y \right)^{2} \right) \right], \label{eq:Cauchy_sum_Oeven} \\
     \mathrm{SO}(2N+1):&& \sum_{\lambda} o^{\operatorname{odd}}_\lambda (X) s_\lambda(Y) 
    & = \prod_{i,j} (1 - x_i y_j)^{-1} \prod_{i \le j} (1 - y_i y_j) \prod_{j}(1-y_{j})^{-1} \nonumber \\ && &
    = \operatorname{PE} \left[\tr X \tr Y \right]
    \operatorname{PE} \left[ \frac{1}{2} \left( -\tr Y^{2}-\left(\tr Y \right)^{2} \right)+\tr Y \right], \label{eq:Cauchy_sum_Oodd} \\
    \mathrm{Sp}(N):&& \sum_{\lambda} sp_\lambda (X) s_\lambda(Y) & = \prod_{i,j} (1 - x_i y_j)^{-1} \prod_{i < j} (1 - y_i y_j) \nonumber \\ && &
    = \operatorname{PE} \left[\tr X \tr Y \right]
    \operatorname{PE} \left[ \frac{1}{2} \left( \tr Y^{2}-\left(\tr Y \right)^{2} \right) \right].
     \label{eq:Cauchy_sum_Sp}
\end{align}
\end{subequations}
The definition of pleythistic exponential is as in Appendix \ref{subsection:weylintegration}. In each of these cases, if the matrices $X$ or $Y$ are elements of the groups $\mathrm{U}(N)$ or $\mathrm{G}(N)$, they have $N$ non-repeated eigenvalues, and the generalized Schur functions vanish for partitions $\lambda$ with depth $\ell(\lambda)=\lambda_{1}^\text{T}>N$. 
For $\ell(\lambda) \le N$ they represent irreducible characters\footnote[2]{There is some subtlety about the irreducibility; see \cite{Koike:1987JA} or \cite{Garcia-Garcia:2019uve} for details.} which are orthonormal under the corresponding group integral,
\begin{subequations} \label{eq:Orthogonality}
\begin{align}
    \int_{\mathrm{U}(N)} \dd{U} s_{\lambda}(U) s_{\mu}(U^{-1}) & = \delta_{\lambda \mu}, \\
    \int_{\mathrm{SO}(2N)} \dd{X} o^{\operatorname{even}}_\lambda(X) o^{\operatorname{even}}_{\mu}(X^{-1}) & = \delta_{\lambda \mu}, \label{eq:Orthogonal_Oeven} \\
    \int_{\mathrm{SO}(2N+1)} \dd{X} o^{\operatorname{odd}}_\lambda(X) o^{\operatorname{odd}}_{\mu}(X^{-1}) & = \delta_{\lambda \mu}, \label{eq:Orthogonal_Oodd} \\
    \int_{\mathrm{Sp}(N)} \dd{X} sp_\lambda(X) sp_{\mu}(X^{-1}) & = \delta_{\lambda \mu}, \label{eq:Orthogonal_Sp}
\end{align}
\end{subequations}
where $\lambda,\mu$ are partitions with $\ell(\lambda),\ell(\mu) \le N$ and the labels $s$, $o$ and $sp$ correspond to the unitary, orthogonal and symplectic characters respectively. 
We note that for $X,Y \in \mathrm{U}(N)$, \eqref{eq:Cauchy_sum_U} just reproduces \eqref{eq:gwworiginalmanyCCschurexpansion}.
The unrestricted Schur sum \eqref{eq:Z_infty} plays the role of normalizer in the subsequent analysis.
We now express the finite $N$ partition function \eqref{eq:gwwgeneralizedmanyCC} in terms of the irreducible characters of the respective groups.
First, for each gauge group we may reparametrize the coupling constants in terms of eigenvalues of matrices $Z$ with eigenvalues $(z_{i})_{i \ge 1}$ and the corresponding Miwa variables $t_{n}$,
\begin{equation}
    N g_n = \frac{2}{n} \tr\left(Z^{n} \right) = \frac{2}{n} t_{n}.
    \label{eq:couplingreparametrizations}
\end{equation}
The matrices $Z$ are infinite dimensional; the Schur polynomial is defined in terms of the infinite array of eigenvalues $(z_{i})_{i \ge 1}$.
\eqref{eq:gwwgeneralizedmanyCC} now decomposes as a product of two plethystic exponentials,
\begin{eqnarray}
\mathcal{Z}_{\mathrm{G}(N)}(\boldsymbol{\upbeta}) &=& \int_{\mathrm{G}(N)} \dd X \, \exp \left( \sum_{n=1}^{\infty} \frac{2}{n} \tr Z^{n} \tr X^{n} \right) \nonumber \\ &=& \int_{\mathrm{G}(N)} \dd X \, \operatorname{PE}\left[ \tr Z \tr X \right]\operatorname{PE}\left[ \tr Z \tr X^{-1} \right], \label{eq:Z_cauchysumbothcases} 
\end{eqnarray}
where we use the reality condition $\tr X= \tr X^{-1}$. 
Now we may use the equations \eqref{eq:Orthogonality} and \eqref{eq:Cauchy_sum} separately in the three cases.

\subsubsection*{Character polynomial expansion for $\mathrm{SO}(2N)$}

The Cauchy sum formula \eqref{eq:Cauchy_sum_Oeven} gives
\begin{equation}
\operatorname{PE} \left[\tr Z \tr X \right]=\operatorname{PE}\left[\frac{1}{2}\left( \tr Z^{2}+(\tr Z)^{2} \right) \right] \sum_{\lambda \mid \ell(\lambda) \le N} s_{\lambda}(Z)o^{\operatorname{even}}_{\lambda}(X). \label{eq:cauchyreexpressionSOeven}
\end{equation}
Plugging this into \eqref{eq:Z_cauchysumbothcases}, and using character reality and \eqref{eq:Orthogonal_Oeven}, we get the random partition representation for the $\mathrm{SO}(2N)$ case,
\begin{eqnarray}
\mathcal{Z}_{\mathrm{SO}(2N)}(\boldsymbol{\upbeta}) &=& \int_{\mathrm{SO}(2N)} \dd X \, \operatorname{PE} \left[ \tr Z^{2}+(\tr Z)^{2} \right] \left( \sum_{\lambda|\ell(\lambda)\le N } s_{\lambda}(Z)o^{\operatorname{even}}_{\lambda}(X) \right) \left( \sum_{\lambda'|\ell(\lambda')\le N } s_{\lambda'}(Z)o^{\operatorname{even}}_{\lambda'}(X^{-1}) \right) \nonumber \\ 
&=& \operatorname{PE} \left[ \tr Z^{2}+(\tr Z)^{2} \right] \sum_{\lambda,\lambda'| \ell(\lambda),\ell(\lambda')\le N} s_{\lambda}(Z)s_{\lambda'}(Z) \int_{\mathrm{SO}(2N)} \dd X \, o^{\operatorname{even}}_{\lambda}(X)o^{\operatorname{even}}_{\lambda'}(X^{-1}) \nonumber \\
&=& \operatorname{PE} \left[ \tr Z^{2}+(\tr Z)^{2} \right] \sum_{\lambda \mid \ell(\lambda)\le N } s_{\lambda}(Z) s_{\lambda}(Z) \nonumber 
\\ &=& \operatorname{PE} \left[ \tr Z^{2} \right] \mathcal{Z}_{\infty}(\boldsymbol{\upbeta},\boldsymbol{0})\mathcal{Z}_{\mathrm{U}(N)}(\boldsymbol{\upbeta},\boldsymbol{0}). \label{eq:randompartitionSONeven}
\end{eqnarray}
The last step in this calculation comes from an application of \eqref{eq:Cauchy_sum_U} and \eqref{eq:Z_infty}, with the specific assignment henceforth of $Z=X=Y$. 

\subsubsection*{Character polynomial expansion for $\mathrm{SO}(2N+1)$}

The $\mathrm{SO}(2N+1)$ case proceeds similarly. The Cauchy sum formula \eqref{eq:Cauchy_sum_Oodd} gives
\begin{equation}
\operatorname{PE} \left[\tr Z \tr X \right]=\operatorname{PE}\left[\frac{1}{2}\left( \tr Z^{2}+(\tr Z)^{2} \right) -\tr Z \right] \sum_{\lambda \mid \ell(\lambda) \le N} s_{\lambda}(Z)o_{\lambda}(X). \label{eq:cauchyreexpressionSOodd}
\end{equation}
Plugging this into \eqref{eq:Z_cauchysumbothcases}, and using character reality and \eqref{eq:Orthogonal_Oodd}, we get the random partition representation for the $\mathrm{SO}(2N+1)$ case,
\begin{eqnarray}
\mathcal{Z}_{\mathrm{SO}(2N+1)}(\boldsymbol{\upbeta}) &=& \int_{\mathrm{SO}(2N+1)} \dd X \, \operatorname{PE} \left[ \tr Z^{2}+(\tr Z)^{2}-2 \tr Z \right] \left( \sum_{\lambda|\ell(\lambda)\le N } s_{\lambda}(Z)o^{\operatorname{odd}}_{\lambda}(X) \right) \left( \sum_{\lambda'|\ell(\lambda')\le N } s_{\lambda'}(Z)o^{\operatorname{odd}}_{\lambda'}(X^{-1}) \right) \nonumber \\ 
&=& \operatorname{PE} \left[ \tr Z^{2}+(\tr Z)^{2}-2 \tr Z \right] \sum_{\lambda,\lambda'|\ell(\lambda),\ell(\lambda')\le N} s_{\lambda}(Z)s_{\lambda'}(Z) \int_{\mathrm{SO}(2N+1)} \dd X \, o^{\operatorname{odd}}_{\lambda}(X)o^{\operatorname{odd}}_{\lambda'}(X^{-1}) \nonumber \\
&=& \operatorname{PE} \left[ \tr Z^{2}+(\tr Z)^{2}-2 \tr Z \right] \sum_{\lambda \mid \ell(\lambda)\le N } s_{\lambda}(Z) s_{\lambda}(Z) \nonumber 
\\ &=& \operatorname{PE} \left[ \tr (Z^{2}-2Z) \right] \mathcal{Z}_{\infty}(\boldsymbol{\upbeta},\boldsymbol{0}) \mathcal{Z}_{\mathrm{U}( N )}(\boldsymbol{\upbeta},\boldsymbol{0}), \label{eq:randompartitionSONodd}
\end{eqnarray}
the last step coming from \eqref{eq:Cauchy_sum_U} and \eqref{eq:Z_infty}. 

\subsubsection*{Character polynomial expansion for $\mathrm{Sp}(N)$}

Finally we find the partition function representation for the $\mathrm{Sp}(N)$ case. 
From \eqref{eq:Cauchy_sum_Sp} we have
\begin{equation}
\operatorname{PE} \left[\tr Z \tr X \right]=\operatorname{PE}\left[\frac{1}{2}\left( -\tr Z^{2}+(\tr Z)^{2} \right) \right] \sum_{\lambda|\ell(\lambda)\le N} s_{\lambda}(Z)sp_{\lambda}(X), \label{eq:cauchyreexpressionSp}
\end{equation}
following which from \eqref{eq:Z_cauchysumbothcases}, character reality and \eqref{eq:Orthogonal_Sp} we get
\begin{eqnarray}
\mathcal{Z}_{\mathrm{Sp}(N)}(\boldsymbol{\upbeta}) &=& \int_{\mathrm{Sp}(N)} \dd X \, \operatorname{PE} \left[ -\tr Z^{2}+(\tr Z)^{2} \right] \left( \sum_{\lambda|\ell(\lambda)\le N} s_{\lambda}(Z)sp_{\lambda}(X) \right) \left( \sum_{\lambda'|\ell(\lambda')\le N} s_{\lambda'}(Z)sp_{\lambda'}(X^{-1}) \right) \nonumber \\ 
&=& \operatorname{PE} \left[ -\tr Z^{2}+(\tr Z)^{2} \right] \sum_{\lambda,\lambda'|\ell(\lambda),\ell(\lambda')\le N} s_{\lambda}(Z)s_{\lambda'}(Z) \int_{\mathrm{Sp}(N)} \dd X \, sp_{\lambda}(X)sp_{\lambda'}(X^{-1}) \nonumber \\
&=& \operatorname{PE} \left[ -\tr Z^{2}+(\tr Z)^{2} \right] \sum_{\lambda|\ell(\lambda)\le N} s_{\lambda}(Z) s_{\lambda}(Z) \nonumber \\
&=& \operatorname{PE} \left[ -\tr Z^{2} \right] \mathcal{Z}_{\infty}(\boldsymbol{\upbeta},\boldsymbol{0}) \mathcal{Z}_{\mathrm{U}(N)}(\boldsymbol{\upbeta},\boldsymbol{0}), \label{eq:randompartitionSpN}
\end{eqnarray}
the last step again coming from \eqref{eq:Cauchy_sum_U} and \eqref{eq:Z_infty}.

\subsection{Free energy}

As with \cref{eq:UNfreeenergy}, we now see from \eqref{eq:randompartitionSONeven}, \eqref{eq:randompartitionSONodd} and \eqref{eq:randompartitionSpN} the free energy in the $\mathrm{SO}$ and $\mathrm{Sp}$ cases has three contributing terms:
\begin{subequations} \label{eq:freeenergyallrandompartition}
\begin{eqnarray} 
\mathcal{F}_{\mathrm{SO}}^{\operatorname{even}}(\boldsymbol{\upbeta}) &=& \lim_{N \rightarrow \infty} \frac{1}{N^{2}} \ln \operatorname{PE}\left[ \tr Z^{2} \right]+ \underbrace{ \lim_{N \rightarrow \infty} \frac{1}{N^{2}} \ln \mathcal{Z}_{\infty}^{2}(\boldsymbol{\upbeta},\boldsymbol{0})}_{=\mathcal{F}_{\mathrm{SO}}^{c,\operatorname{even}}(\boldsymbol{\upbeta})}+\underbrace{\lim_{N \rightarrow \infty} \frac{1}{N^{2}} \ln \left( \frac{Z_{\mathrm{U}(N)}(\boldsymbol{\upbeta},\boldsymbol{0})}{\mathcal{Z}_{\infty}(\boldsymbol{\upbeta},\boldsymbol{0})} \right)}_{\mathcal{F}_{\mathrm{SO}}^{f,\operatorname{even}}(\boldsymbol{\upbeta})}, \label{eq:freeenergySOeven,randompartition} \\
\mathcal{F}_{\mathrm{SO}}^{\operatorname{odd}}(\boldsymbol{\upbeta}) &=& \lim_{N \rightarrow \infty} \frac{1}{N^{2}} \ln \operatorname{PE}\left[ \tr (Z^{2}-2Z) \right]+ \underbrace{ \lim_{N \rightarrow \infty} \frac{1}{N^{2}} \ln \mathcal{Z}_{\infty}^{2}(\boldsymbol{\upbeta},\boldsymbol{0})}_{=\mathcal{F}_{\mathrm{SO}}^{c,\operatorname{odd}}(\boldsymbol{\upbeta})}+\underbrace{\lim_{N \rightarrow \infty} \frac{1}{N^{2}} \ln \left( \frac{Z_{\mathrm{U}(N)}(\boldsymbol{\upbeta},\boldsymbol{0})}{\mathcal{Z}_{\infty}(\boldsymbol{\upbeta},\boldsymbol{0})} \right)}_{\mathcal{F}_{\mathrm{SO}}^{f,\operatorname{odd}}(\boldsymbol{\upbeta})}, \label{eq:freeenergySOodd,randompartition} \\
\mathcal{F}_{\mathrm{Sp}}(\boldsymbol{\upbeta}) &=& \lim_{N \rightarrow \infty} \frac{1}{N^{2}} \ln \operatorname{PE}\left[ -\tr Z^{2} \right]+ \underbrace{ \lim_{N \rightarrow \infty} \frac{1}{N^{2}} \ln \mathcal{Z}_{\infty}^{2}(\boldsymbol{\upbeta},\boldsymbol{0})}_{=\mathcal{F}_{\mathrm{Sp}}^{c}(\boldsymbol{\upbeta})}+\underbrace{\lim_{N \rightarrow \infty} \frac{1}{N^{2}} \ln \left( \frac{Z_{\mathrm{U}(N)}(\boldsymbol{\upbeta},\boldsymbol{0})}{\mathcal{Z}_{\infty}(\boldsymbol{\upbeta},\boldsymbol{0})} \right)}_{\mathcal{F}_{\mathrm{Sp}}^{f}(\boldsymbol{\upbeta})}. \label{eq:freeenergySpNrandompartition}
\end{eqnarray}
\end{subequations}
We assume the limits are well-defined. 
It is immediate to notice from a comparison with \eqref{eq:UNfreeenergy} that the continuum component in all the cases is functionally twice that in the $\mathrm{U}(N)$ case with switched-off imaginary couplings, and the fluctuation components are identical:
\begin{subequations}
\begin{align}
\mathcal{F}^{c,\operatorname{even}}_{\mathrm{SO}} (\boldsymbol{\upbeta}) &= \mathcal{F}^{c,\operatorname{odd}}_{\mathrm{SO}}(\boldsymbol{\upbeta})=\mathcal{F}^{c}_{\mathrm{Sp}}(\boldsymbol{\upbeta})=2\mathcal{F}^{c}_{\mathrm{U}}(\boldsymbol{\upbeta},\boldsymbol{0}), \label{eq:continuumtwice} \\   
\mathcal{F}^{f,\operatorname{even}}_{\mathrm{SO}}(\boldsymbol{\upbeta}) &= \mathcal{F}^{f,\operatorname{odd}}_{\mathrm{SO}}(\boldsymbol{\upbeta})= \mathcal{F}^{f}_{\mathrm{Sp}}(\boldsymbol{\upbeta})=\mathcal{F}^{f}_{\mathrm{U}}(\boldsymbol{\upbeta},\boldsymbol{0}). \label{eq:fluctuationequal}   
\end{align}
\end{subequations}
Hence both the continuum and fluctuation parts may be treated identically, in analysis of qualitative behaviour, to the $\mathrm{U}(N)$ treatment in \cite{2021taroalisecond}. 
Just like for the $\mathrm{U}(N)$ model, a straightforward calculation using \eqref{eq:Cauchy_sum_U} and \eqref{eq:couplingreparametrizations} shows that the continuum part is the ungapped phase free energy \eqref{eq:freeenergy}. 

\subsubsection*{Asymptotics of the residual term}

We are left with the analysis of the first term in the \eqref{eq:freeenergyallrandompartition}, which is different in the three cases. 
Denoting it to be $\mathcal{R}^{\operatorname{even}}_{\mathrm{SO}}(\boldsymbol{\upbeta}),\mathcal{R}^{\operatorname{odd}}_{\mathrm{SO}}(\boldsymbol{\upbeta}),\mathcal{R}_{\mathrm{Sp}}(\boldsymbol{\upbeta})$ respectively, we note that they have the following explicit expression in terms of the coupling constants, using \eqref{eq:Cauchy_sum_U} and \eqref{eq:couplingreparametrizations}: 
\begin{subequations}
\begin{eqnarray}
\mathcal{R}^{\operatorname{even}}_{\mathrm{SO}}(\boldsymbol{\upbeta}) &=& \lim_{N \rightarrow \infty} \frac{1}{N} \sum_{n \ge 1} \frac{\beta_{2n}}{2n}, \label{eq:extra_term_freeenergySOeven} \\
\mathcal{R}^{\operatorname{odd}}_{\mathrm{SO}}(\boldsymbol{\upbeta}) &=& -\lim_{N \rightarrow \infty} \frac{1}{N} \sum_{n \ge 1} \frac{\beta_{2n-1}}{2n-1}, \label{eq:extra_term_freeenergySOodd} \\
\mathcal{R}_{\mathrm{Sp}}(\boldsymbol{\upbeta}) &=& -\lim_{N \rightarrow \infty} \frac{1}{N} \sum_{n \ge 1} \frac{\beta_{2n}}{2n}, \label{eq:extra_term_freeenergySp}.   
\end{eqnarray}
\end{subequations}
Hence they asymptotically go as $ \mathcal{O}\left( \frac{1}{N} \right)$ and do not take part in the phase transitions. 
In fact, they may be considered as subleading $\mathcal{O}\left( \frac{1}{N} \right)$ contributions to the continuum component. 
We note that we corroborate the expressions $\exp\left(N^{2}\mathcal{R}^{\operatorname{even}}_{\mathrm{SO}}(\boldsymbol{\upbeta})\right)\mathcal{Z}_{\infty}^{2} (\boldsymbol{\upbeta})$, $\exp \left(N^{2} \mathcal{R}^{\operatorname{odd}}_{\mathrm{SO}}(\boldsymbol{\upbeta}) \right)\mathcal{Z}_{\infty}^{2} (\boldsymbol{\upbeta})$ and $\exp \left(N^{2}\mathcal{R}_{\mathrm{Sp}}(\boldsymbol{\upbeta})\right)\mathcal{Z}_{\infty}^{2} (\boldsymbol{\upbeta})$\footnote[1]{In each case, the $N^{2}$ factor preceding the $\mathcal{R}$-term is to be understood as pre-limit.} 
respectively with the results obtained due to the Szeg\"o--Johansson theorem \cite{Johansson1997} in \cite[Appendix B]{Garcia-Garcia:2019uve}, where in Eqs. (77) to (79) we have the correspondence $V_{k}= \frac{N \beta_{k}}{k}$.
In the context of the Coulomb gas formalism described in \cref{sec:generalformalism}, an interpretation of these $\mathcal{R}$-terms may be given by the sub-leading $\Xi$- and single-variable action terms which we discarded in \eqref{eq:gwwgeneralizedmanyCCgeneralformlargeN} and \eqref{eq:finiteNeffactionderivative}.
The phase transition dynamics are entirely due to the fluctuation free energy $\mathcal{F}_{\mathrm{G}}^{f}=\mathcal{F}_{\mathrm{U}}^{f}$, with edge asymptotics described by \eqref{eq:freeenergyU(N)edgelargeN}.
We note that the correspondence between the $\beta_{n}$ and the $t_{n}$ as defined in \eqref{eq:couplingreparametrizations} is the same as for the unitary model, i.e. \eqref{eq:miwa_variablesUN}. 
This means that all the notation used in the review in \cref{subsec:reviewUNrandompart} can be directly interpreted without changes.

\subsection*{Acknowledgements}
This work was supported by ``Investissements d'Avenir'' program, Project ISITE-BFC (No.~ANR-15-IDEX-0003), EIPHI Graduate School (No.~ANR-17-EURE-0002), and Bourgogne-Franche-Comté region. SP acknowledges discussions with Ali Zahabi. We thank the reviewer for several useful suggestions and remarks.

\subsection*{Erratum}
Since the publication of this work we have noticed a minor error we would like to correct in this preprint version.
The odd-orthogonal Cauchy formula \eqref{eq:Cauchy_sum_Oodd} should in fact be
\begin{align}
    \mathrm{SO}(2N+1):&& \sum_{\lambda} o^{\operatorname{odd}}_\lambda (X) s_\lambda(Y) 
    & = \prod_{i,j} (1 - x_i y_j)^{-1} \prod_{i \le j} (1 - y_i y_j) \prod_{j}(1-y_{j})^{-1} \nonumber \\ && &
    = \operatorname{PE} \left[\tr X \tr Y \right]
    \operatorname{PE} \left[ \frac{1}{2} \left( -\tr Y^{2}-\left(\tr Y \right)^{2} \right) \right], \label{eq:Cauchy_sum_Oodd_revised} 
\end{align}
i.e. the plethystic exponential form is formally identical to that of the even orthogonal Cauchy formula \eqref{eq:Cauchy_sum_Oeven}. 
This is because the arguments of the generalized Schur function $o^{\operatorname{odd}}_{\lambda}$ are the non-trivial eigenvalues $x_{i}$ of a matrix $X \in \mathrm{SO}(2N+1)$ only; the eigenvalue of $1$ is not an argument (the footnote is erroneous in this regard).
Hence in the second equality of \eqref{eq:Cauchy_sum_Oodd}, we should actually have $\prod_{i,j}(1-x_{i}y_{j})^{-1}\prod_{j}(1-y_{j})^{-1}=\operatorname{PE}\left[\tr X \tr Y\right]$.
All contributions from the $\operatorname{PE} \left[\tr Y \right]$ factor should not be present in subsequent calculations for the odd orthogonal case, and \eqref{eq:cauchyreexpressionSOodd} and \eqref{eq:randompartitionSONodd} should be identical to \eqref{eq:cauchyreexpressionSOeven} and \eqref{eq:randompartitionSONeven} respectively. 
Eventually, the subleading term in the free energy for the odd orthogonal case, \eqref{eq:extra_term_freeenergySOodd}, should be identical to that for the even orthogonal case, \eqref{eq:extra_term_freeenergySOeven}, i.e.
\begin{eqnarray}
    \mathcal{R}^{\operatorname{odd}}_{\mathrm{SO}}(\boldsymbol{\upbeta}) &=& \lim_{N \rightarrow \infty} \frac{1}{N} \sum_{n \ge 1} \frac{\beta_{2n}}{2n}. \label{eq:extra_term_freeenergySOodd_revised} \\
\end{eqnarray}
In the formulae ~\eqref{eq:Cauchy_sum_U} through ~\eqref{eq:Cauchy_sum_Sp}, we make it clear that the matrix argument $X$ is assumed to belong to the respective compact classical group.
The matrix argument $Y$ could be unrestricted.

The correction \eqref{eq:extra_term_freeenergySOodd_revised}, as it is subleading at the $\mathcal{O}\left(\frac{1}{N}\right)$ level, does not change the phase spectrum interpretation results of the paper.
We also continue to have corroboration with the Szeg\"o--Johansson theorem \cite{Johansson1997}.
However we lose the ability to distinguish odd orthogonal asymptotics from even orthogonal asymptotics at the $\mathcal{O}(\frac{1}{N})$ level.

\appendix

\section{Weyl integration formula}
\label{subsection:weylintegration}

Let $G$ be a compact Lie group with maximal torus $T$. Assume that both have normalized Haar measures defined on them. Then for any continuous complex function $f$ on $G$, the Weyl integral formula may be written as 
\begin{equation}
\int_G \dd g \, f(g) = \frac{1}{|W|} \int_{T} \dd t \left[ \det \left(I_{G/T}-\operatorname{Ad}_{G/T}(t^{-1}) \right) \int_{G} \dd g \, f(gtg^{-1}) \right]. \label{eq:weylintegrationformulamain}
\end{equation}
Here $W$ is the Weyl group of $G$, and the Vandermonde determinant $J(t)=\det \left(I_{G/T}-\operatorname{Ad}_{G/T}(t^{-1})\right)$ which appears as a Jabobian for the change of measure depends on the elements of the maximal torus. The function $f$ in the context of our work is a class function composed of characters. We refer to standard texts, e.g. \cite{Adams1983-vh,fultonharris} for further details on the terms appearing in this formula. 

\subsection*{Vandermonde determinant}
\label{subsubsection:vandermonde}

The Vandermonde determinant $J(t)$ is related to the root system of $G$. 
It may be written in terms of the torus parameters as \cite{fultonharris}
\begin{equation}
J(t)=\frac{1}{\left| \operatorname{PE}\left[ \sum_{\alpha} \prod_{k}a_{k}^{\alpha_{k}} \right] \right|}= \prod_{\alpha}   \left|1-\prod_{k}a_{k}^{\alpha_{k}} \right|, \label{eq:vandermondePE}
\end{equation}
where $\alpha = (\alpha_k)$ is the set of non-zero roots of $G$ and the $a_{k}$ parameterize the maximal torus. 
The argument $t \in T$, an element of the maximal torus, may be described as a matrix $\operatorname{diag}(a_{k})$.
The operation of plethystic exponentiation is defined as 
\begin{equation}
\operatorname{PE}[f(x_{i})]= \exp \left( \sum_{n=1}^{\infty} \frac{1}{n}f(x_{i}^{n})\right). \label{eq:plethysticdefn}
\end{equation}

For the gauge groups relevant to this paper, the Vandermonde determinant may be written with the angular parametrization $a_{k}=e^{i\phi_k}$ as follows (see \cite{fultonharris}):
\begin{subequations} \label{eq:vandermonde}
\begin{align}
\mathrm{U}(N): & \;\;\; \prod_{\substack{k,l=1 \\ k \ne l}}^{N} \left|e^{i\phi_{k}}-e^{i\phi_{l}} \right|, \label{eq:vandermonde1} \\
\mathrm{SO}(2N): & \;\;\; 2 \prod_{\substack{k,l=1 \\ k \ne l}}^{N} \left|e^{\frac{i}{2}(\phi_{k}+\phi_{l})}-e^{-\frac{i}{2}(\phi_{k}+\phi_{l})} \right|\left|e^{\frac{i}{2}(\phi_{k}-\phi_{l})}-e^{-\frac{i}{2}(\phi_{k}-\phi_{l})} \right|, \label{eq:vandermonde2} \\
\mathrm{SO}(2N+1): & \;\;\; \prod_{\substack{k,l=1 \\ k \ne l}}^{N} \left|e^{\frac{i}{2}(\phi_{k}+\phi_{l})}-e^{-\frac{i}{2}(\phi_{k}+\phi_{l})} \right|\left|e^{\frac{i}{2}(\phi_{k}-\phi_{l})}-e^{-\frac{i}{2}(\phi_{k}-\phi_{l})} \right| \prod_{k=1}^{N}\left|e^{\frac{i\phi_{k}}{2}}-e^{\frac{-i\phi_{k}}{2}}\right|^{2}, \label{eq:vandermonde3} \\
\mathrm{Sp}(N): & \prod_{\substack{k,l=1 \\ k \ne l}}^{N} \left|e^{\frac{i}{2}(\phi_{k}+\phi_{l})}-e^{-\frac{i}{2}(\phi_{k}+\phi_{l})} \right|\left|e^{\frac{i}{2}(\phi_{k}-\phi_{l})}-e^{-\frac{i}{2}(\phi_{k}-\phi_{l})} \right| \prod_{k=1}^{N}\left|e^{i\phi_{k}}-e^{-i\phi_{k}}\right|^{2}. \label{vandermonde4}
\end{align}
\end{subequations}.

\subsection*{Effective action parameters}
\label{subsubsection:fredholm}
The integral equation kernels $\Delta(\phi,\varphi)$ may then be obtained from the equations \eqref{eq:vandermonde}  by a re-expression of the Vandermonde determinant using logarithms and symmetries of the sums. 
More precisely, we take the double-index product sectors of the equations~\eqref{eq:vandermonde}, while ignoring the numerical pre-factor of $2$ in the $\mathrm{SO}(2N)$ case, and the single-index product factors in the $\mathrm{SO}(2N+1)$ and $\mathrm{Sp}(N)$ cases.
In the exponent they become sums over logarithms as follows:
\begin{subequations} \label{eq:buildingDelta}
\begin{align}
\mathrm{U}(N): & \;\;\; \ln \left( \prod_{\substack{k,l=1 \\ k \ne l}}^{N} \left|e^{i\phi_{k}}-e^{i\phi_{l}} \right| \right) = \frac{1}2 \sum_{\substack{k,l=1 \\k \ne l}}^{N} \ln \left(4 \sin^{2}\left( \frac{\phi_{k}-\phi_{l}}{2} \right) \right), \label{eq:buildingDeltaUN} \\
\mathrm{SO}(2N),\mathrm{SO}(2N+1),\mathrm{Sp}(N): & \;\;\; \ln \left( \prod_{\substack{k,l=1 \\ k \ne l}}^{N} \left|e^{\frac{i}{2}(\phi_{k}+\phi_{l})}-e^{-\frac{i}{2}(\phi_{k}+\phi_{l})} \right|\left|e^{\frac{i}{2}(\phi_{k}-\phi_{l})}-e^{-\frac{i}{2}(\phi_{k}-\phi_{l})} \right| \right) \nonumber \\ & \;\;\; = \frac{1}2 \sum_{\substack{k,l=1 \\k \ne l}}^{N} \left[ \ln \left(4 \sin^{2}\left( \frac{\phi_{k}+\phi_{l}}{2} \right) \right)+ \ln \left(4 \sin^{2}\left( \frac{\phi_{k}-\phi_{l}}{2} \right) \right)\right]. \label{eq:buildingDeltaSOSp}
\end{align}
\end{subequations}
Hence we get the expressions $\Delta(\phi,\varphi)$ for the three cases:
\begin{subequations} \label{eq:fredholm}
\begin{align}
\mathrm{U}(N): & \;\;\;  \frac{1}{2} \ln \left (4 \sin^{2} \left( \frac{\phi-\varphi}{2} \right) \right), \label{eq:fredholm1} \\
\mathrm{SO}(2N),\mathrm{SO}(2N+1),\mathrm{Sp}(N): & \;\;\; \frac{1}{2} \left[ \ln \left (4 \sin^{2} \left( \frac{\phi+\varphi}{2} \right)\right)+\ln \left (4 \sin^{2} \left( \frac{\phi-\varphi}{2}\right) \right) \right]. \label{eq:fredholm234}
\end{align}
\end{subequations}
For completeness we also include the $\mathrm{U}(N)$ kernel corresponding to the model \eqref{eq:gwworiginalmanyCC}, which we do not treat in detail in this paper. The single-index or numerical product factors similarly also become sums over logarithms in the exponent:
\begin{subequations} \label{eq:buildingXi}
\begin{align}
\mathrm{SO}(2N): & \;\;\; \ln 2 = \frac{1}{N} \sum_{k=1}^{N} \ln 2, \label{eq:buildingXiSO2N} \\
\mathrm{SO}(2N+1): & \;\;\; \ln \left( \prod_{k=1}^{N}\left|e^{\frac{i\phi_{k}}{2}}-e^{\frac{-i\phi_{k}}{2}}\right|^{2} \right) = \sum_{k=1}^{N} \ln \left( 4 \sin^{2} \left ( \frac{\phi_{k}}{2} \right) \right), \label{eq:buildingXiSO2N+1} \\
\mathrm{Sp}(N): & \;\;\; \ln \left(\prod_{k=1}^{N}\left|e^{i\phi_{k}}-e^{-i\phi_{k}}\right|^{2}\right)=\sum_{k=1}^{N}\ln \left ( 4 \sin^{2}  \phi_{k} \right). \label{eq:buildingXiSpN}
\end{align}
\end{subequations}
Hence we obtain the respective subleading contributions $\Xi(\phi)$:
\begin{subequations} \label{eq:xi}
\begin{align}
\mathrm{SO}(2N): & \;\;\; \frac{1}{N} \ln 2, \label{eq:xiSO2N} \\
\mathrm{SO}(2N+1): & \;\;\; \ln \left( 4 \sin^{2} \left ( \frac{\phi}{2} \right) \right), \label{eq:xiSO2N+1} \\
\mathrm{Sp}(N): & \;\;\; \ln \left ( 4 \sin^{2}  \phi \right). \label{eq:xiSpN}
\end{align}
\end{subequations}

\section{Maximal tori}
\label{subsection:maximaltori}

All maximal tori in a Lie group are conjugate to one another and formally equivalent in choice in the Weyl integration formula \eqref{eq:weylintegrationformulamain}. For convenience we choose the following canonical maximal tori in our analysis: 
\begin{subequations}
\begin{itemize}
\item[a)] For $\mathrm{U}(N)$, we take the set of $N \times N$ diagonal matrices, 

\begin{equation}
\textrm{diag}\left(e^{i\phi_{1}},\ldots,e^{i\phi_{N}} \right). \label{eq:UNtorusmatrices}
\end{equation}

\item[b)] For $\mathrm{SO}(2N)$, we take the set of $2N \times 2N$ block-diagonal matrices in $2 \times 2$ blocks of $\mathrm{SO}(2)$ rotation matrices,

\begin{equation}
\textrm{diag}\left( \begin{pmatrix}
\cos \phi_{1} & - \sin \phi_{1} \\ \sin \phi_{1} & \cos \phi_{1} 
\end{pmatrix},\ldots,\begin{pmatrix}
\cos \phi_{N} & - \sin \phi_{N} \\ \sin \phi_{N} & \cos \phi_{N} 
\end{pmatrix} \right). \label{eq:So2Ntorusmatrices}
\end{equation}

\item[c)] For $\mathrm{SO}(2N+1)$, the form of the elements is nearly the same as for $\mathrm{SO}(2N)$ but with an additional solitary diagonal entry of $1$, conventionally put at the upper left, i.e. the set of $(2N+1) \times (2N+1)$ matrices

\begin{equation}
\textrm{diag}\left(1,\begin{pmatrix}
\cos \phi_{1} & - \sin\phi_{1} \\ \sin \phi_{1} & \cos \phi_{1} 
\end{pmatrix},\ldots,\begin{pmatrix}
\cos \phi_{N} & - \sin \phi_{N} \\ \sin \phi_{N} & \cos \phi_{N} 
\end{pmatrix} \right). \label{eq:So2N+1torusmatrices}
\end{equation}

\item[d)] For $\mathrm{Sp}(N)$ we take the set of $2N \times 2N$ diagonal matrices 
\begin{equation}
\operatorname{diag}(e^{\pm i \phi_1},\ldots,e^{\pm i \phi_N}), \label{eq:SpNtorusmatrices}
\end{equation}
i.e. two copies of the canonical $\mathrm{U}(N)$ tori. 

\end{itemize}
\end{subequations}

We note that the $\mathrm{SO}(N)$ tori are $N$-dimensional by the identification $\mathrm{U}(1) \cong \mathrm{SO}(2)$. 
Further, we remark that these decompositions show that the non-trivial eigenvalues of the real matrices come in complex conjugate pairs -- with the odd orthogonal case having an extra trivial eigenvalue of $1$. 

\subsection*{Single-variable action}
\label{subsubsection:singlevariableaction}

Using the maximal tori \eqref{eq:UNtorusmatrices} through \eqref{eq:SpNtorusmatrices}, we derive the single-variable action as defined in \eqref{eq:disceteNactions}. For completeness we also mention the $\mathrm{U}(N)$ action corresponding to \eqref{eq:gwworiginalmanyCC}. The actions with their usual real parametrizations are:

\begin{subequations} \label{eq:Sallcases}
\begin{align}
\mathrm{U}(N): & \;\;\; \sum_{n \ge 1} \frac{1}{n} \left( \beta_{n} \cos n \phi+ \gamma_{n} \sin n \phi \right), \label{eq:single_variable_actionU(N)} \\
\mathrm{SO}(2N),\mathrm{Sp}(N): & \;\;\;  \sum_{n \ge 1} \frac{2\beta_{n}}{n} \cos n \phi, \label{eq:single_variable_actionSO(2N),Sp(N)} \\ 
\mathrm{SO}(2N+1): & \;\;\; \sum_{n \ge 1} \frac{2\beta_{n}}{n} \left( \cos n \phi+\frac{1}{2N} \right). \label{eq:single_variable_actionSO(2N+1)} 
\end{align}
\end{subequations}
The $\mathcal{O}\left(N^{-1}\right)$ term in the $\mathrm{SO}(2N+1)$ case comes from the solitary matrix entry of $1$ and can be ignored, hence the expressions \eqref{eq:single_variable_actionSO(2N),Sp(N)} and \eqref{eq:single_variable_actionSO(2N+1)} may be considered identical. We note that this particular action is even.

\section{Mathematical identities}
\subsection*{Fourier series}
\label{subsubsec:fourierseries}
The Fourier series for $\cot\left( \frac{x}{2} \right)$ may be derived as follows. We consider the formal expansion (see the discussion following \eqref{eq:finiteNeffactionderivativemodified} for the notation)
\begin{equation}
\sum_{n=1}^{\infty} e^{in(x+i0)}= \frac{e^{i(x+i0)}}{1-e^{i(x+i0)}}=\frac{i}{2} \cot \left(\frac{x}{2}+i0 \right)- \frac{1}{2}. \label{eq:proofexpansioncot2}
\end{equation}
Equating imaginary parts of \eqref{eq:proofexpansioncot2} and performing the regularization gives the required result
\begin{equation}
\cot \left( \frac{x}{2} \right) = 2 \sum_{n=1}^{\infty} \sin nx. \label{eq:fourierseriescot}
\end{equation}
\subsection*{Plemelj formula}
\label{subsubsection:plemelj}
For the discrete resolvent \eqref{eq:resolvent} we define the even discrete probability measure based on the entries $(\phi_{l})_{1 \le l \le N}$,
\begin{equation}
\rho(\phi)= \frac{1}{2N} \sum_{l=1}^{N} \left[ \delta(\phi-\phi_{l})+\delta (\phi+\phi_{l}) \right]. \label{eq:discreteprobability}
\end{equation}
This is a valid probability measure for the set of singularities $\pm \phi_{l}$ of \eqref{eq:resolvent}.
It is even, and this evenness is preserved in the large $N$ limit. For any interval $I \subset [-\pi,\pi)$ let us define the multisets $A_{I}=\{\phi_{l} \in I \}$, $B_{I}=\{-\phi_{l} \in I \}$, then we have
\begin{equation}
\int_{I} \dd \phi \, \rho(\phi)= \frac{|A_{I}|+|B_{I}|}{2N}. \label{eq:integraldiscreteprobabilityI}    
\end{equation}
By contour integration arguments, we have
\begin{equation}
\int_{I} \dd \phi \, W(\phi-i0) - \int_{I} \dd \phi \, W(\phi+i0)  = \sum_{\phi_{l} \in A_{I}} \oint_{C(\phi_{l},R \rightarrow 0)} \dd \phi \, W(\phi)+ \sum_{-\phi_{l} \in B_{I}} \oint_{C(-\phi_{l},R \rightarrow 0)} \dd \phi \, W(\phi), \label{eq:sumofresidues}    
\end{equation}
where we have written the difference of the two integrals as a sum over vanishing anticlockwise contours around the relevant residues. From the form of the resolvent \eqref{eq:resolvent}, we see that only particular terms contribute to this sum: 
\begin{equation}
\int_{I} \dd \phi \, W(\phi-i0) - \int_{I} \dd \phi \, W(\phi+i0)  = \frac{1}{N} \sum_{\phi_{l}\in A_{I}} \oint_{C(\phi_{l},R \rightarrow 0)} \dd \phi \, \cot \left( \frac{\phi-\phi_{l}}{2} \right)+ \frac{1}{N} \sum_{-\phi_{l} \in B_{I}} \oint_{C(-\phi_{l},R \rightarrow 0)} \dd \phi \, \cot \left( \frac{\phi+\phi_{l}}{2} \right). \label{eq:sumofresiduesnew}    
\end{equation}
These contour integrals may be evaluated to give
\begin{equation}
\int_{I} \dd \phi \, W(\phi-i0) - \int_{I} \dd \phi \, W(\phi+i0) = \frac{4\pi i}{N} \left(|A_{I}|+|B_{I}| \right). \label{eq:plemejlformula}  
\end{equation}
Comparing with \eqref{eq:integraldiscreteprobabilityI} and taking into consideration the arbitrariness of $I$ gives the result \eqref{eq:densityrecoveryresolvent} in the continuum limit.

\subsection*{One-gap probability distribution}
\label{subsubsec:proofofc}

Using the notation introduced at the end of \cref{subsection:gappedphase}, let us analyze the definite integral 
\begin{equation}
I(c)=\frac{1}{\pi} \int_{0}^{\alpha(c)} \sqrt{-\beta^{2}\sin^{2}\phi+2\beta \cos \phi+c}, \label{eq:definiteintegralasc}     
\end{equation}
where we have used the symmetry of the function to take just the positive part of the domain, and $\alpha(c)$ is the smallest real number in $[0,\pi)$ such that $f(\phi;c)=-\beta^{2}\sin^{2}\phi+2\beta \cos \phi+c=0$. 
A straightforward analysis of the extrema of $f(\phi;c)$ in $[0,\pi]$ shows that for exactly $-2\beta<c <1+\beta^{2}$, $f(\phi;c)>0$ in $[0,\alpha(c))$, and strictly decreasing in $(0,\alpha(c)]$.   
This determines the condition for a gap to appear. 
Assuming $c$ to lie in this range, we observe that $I(c)$ as a function of $c$ is strictly increasing on $(-2\beta,1+\beta^{2})$. 
Hence there will be, if at all, a unique $c$ such that $I(c)=1$. 
We now evaluate $I(2\beta)$ using the substitution $z=\sin\left(\frac{\phi}{2}\right)$ and $\alpha(2\beta)=2 \arcsin{ \frac{1}{\sqrt{\beta}}}$:
\begin{eqnarray}
I(2\beta) &=& \frac{2 \beta}{\pi} \int_{0}^{\alpha(2\beta)} \dd \phi \, \sqrt{\frac{1}{\beta}-\sin^{2}\left(\frac{\phi}{2}\right)} \nonumber
\\
&=& \frac{4 \beta}{\pi} \int_{0}^{\frac{1}{\sqrt{\beta}}} \dd z \, \sqrt{\frac{1}{\beta}-z^{2}} \nonumber \\
&=& \frac{4 \beta}{\pi} \left[ \frac{z}{2} \sqrt{\frac{1}{\beta}-z^{2}}+\frac{1}{2\beta} \arcsin \sqrt{\beta}z \right]_{0}^{\frac{1}{\sqrt{\beta}}} \nonumber \\
&=& 1. \label{eq:normalizationshown}
\end{eqnarray}
Hence $c=2\beta$ fixes the normalization and is the unique such $c$.

\subsection*{Reality of compact classical group characters}
\label{subsubsec:realityofcharacters}

We show that, for $X \in \mathrm{G}(N)$ as defined by \eqref{eq:listofgaugegroups}, we have $\tr X = \tr X^{-1} \in \mathbb{R}$. 

For the special orthogonal case, the reality is immediate from the real nature of the matrices, and equality is immediate from the observation that $X^\text{T}=X^{-1}$ and $\tr X=\tr X^\text{T}$.

For the symplectic case, let us consider the elements $X,X^{-1}$ of $\mathrm{Sp}(N)=\mathrm{Sp}(2N,\mathbb{C}) \cap \mathrm{U}(2N)$ in the canonical representation of $2N \times 2N$ block-diagonal matrices, i.e.
\begin{equation}
X= \begin{pmatrix} A & B \\ C & D \end{pmatrix}, \;\;\; \Omega=\begin{pmatrix}
0 & 1 \\ -1 & 0    
\end{pmatrix}, \;\;\; X^{-1}=-\Omega X^\text{T} \Omega= \begin{pmatrix}
D^\text{T} & -B^\text{T} \\ -C^\text{T} & A^\text{T}
\end{pmatrix}.  \label{eq:decompositionofSpNmatrixall}
\end{equation}
Here all the entries $A,B,C,D,0,\pm 1$ are $N \times N$ complex matrices, and the expression for the inverse is derived from the symplectic condition $X^{T}\Omega X=\Omega$, or equivalently, $X \Omega X^{T}=\Omega$. 
The equality condition follows since $\tr X^{-1}= -\tr \Omega X^\text{T} \Omega = - \tr \Omega^{2} X^\text{T}= \tr X$. 
To obtain the reality condition, $X \in \mathrm{U}(2N)$ so we must have $X^{\dag}=X^{-1}$, so from the block decompositions we get $A^{\dag}=D^\text{T}, B^{\dag}=-C^\text{T}$. 
This means 
\begin{equation}
\tr X= \tr A+ \tr D = \tr A+\tr A^{*} \in \mathbb{R}. \label{eq:tracesequalsymplectic}
\end{equation}

\bibliographystyle{utphys}
\bibliography{refs}
\end{document}